\documentclass[a4paper]{article} 
\usepackage{epsfig}
\usepackage{amsmath}
\usepackage[left=2cm,top=2cm,bottom=3cm]{geometry}

\makeatletter 
\@addtoreset{equation}{section} 
\makeatother 
 
\newcommand{\be}{\begin{equation}}
\newcommand{\ee}{\end{equation}}
\newcommand{\bea}{\begin{eqnarray}}
\newcommand{\eea}{\end{eqnarray}}

\def\p{\partial }

\def\ep{\epsilon}

\def\ov{\over  }

\begin{document}

\centerline{\LARGE{\bf  {Sommerfeld enhancement: }}} 
\vskip0.3cm
 \centerline{\LARGE{\bf  {general results from field theory diagrams}}} 
\vskip 0.3cm
\centerline{\bf Roberto Iengo \footnote{iengo@sissa.it}}

\vskip .3cm
\centerline{\it  International School for Advanced Studies (SISSA)}
\centerline{\it Via Beirut 2-4, I-34013 Trieste, Italy} 
\centerline{\it  INFN, Sezione di Trieste}

\vskip0.3cm

{\bf Abstract.} Assuming that two incoming annihilating particles interact by  exchanging
a generally massive attractive vector (or scalar) boson,
we find, by taking the non-relativistic limit of the field theory ladder
diagrams,  that the complete annihilation amplitude $A$ is equal to: the convolution
of a solution of the Schroedinger equation (including the attractive potential) 
with the Fourier transform of the bare (i.e. ignoring the attraction) annihilation amplitude $A_0$. 
The main novelty is that $A_0$ can be completely arbitrary.
For a Coulomb potential we find analytically the enhancement for the l-partial-wave cross-section,
e.g. the P wave enhancement  $2\pi(\alpha/ v)^3$ ($v$ relative velocity), for a Yukawa potential 
we describe a simple algorithm and give numerical results showing an important P wave enhancement with a resonant pattern.


\section{Summary and results}

Here we discuss general results for the Sommerfeld enhancement, which could be an important ingredient for the interpretation of astrophysical data to signal dark matter annihilation processes\cite{Hisa},\cite{ Stru},\cite{Mar},\cite{Stru2}, \cite{ArkHam},\cite{Lat},\cite{Cli},\cite{Rob}.

We assume that two incoming particles attract each other by exchanging repeatedly  a  massive vector boson, before undergoing some annihilation reaction. We will be interested in the non-relativistic limit of this process: in this limit the exchange of a scalar boson
 would give the same result, although to be definite we will continue to refer to a vector boson.
 
 Also, in general one has to consider a non-abelian process, corresponding to a matrix-valued interaction. 
 In the analysis below we assume that 
that  the matrix can be diagonalized and  that  there is a boson of definite mass mediating the attraction 
for the relevant eigenstates of the  two particles. We consider therefore such a definite eigenstate.

Let $\vec p$ be the CM momentum 
of the incoming particles and $m$ their mass.
We call $A_0(\vec p)$ the bare  (i.e. neglecting the vector field  attraction) annihilation
amplitude and $A(\vec p)$ the 
complete (including the effect of the attraction) amplitude.  

Our general result is a generalization of the enhancement formula of ref.\cite{Hisa}.
 Our treatment focuses on the relation 
 between $A(\vec p)$ and $A_0(\vec p)$ for a totally generic $A_0(\vec p)$,
 rather than taking for instance a constant $A_0$ (i.e. its
 Fourier transform $\sim\delta(r)$).
 This allows us in particular to consider the specific case of  
 partial waves higher than S.  

Our method is also similar to the one of ref.\cite{Hisa}, even if the path we follow is not quite the same.
In a sense it is more direct, since for simplicity we focus on the essential points, leaving off the details of
the dark matter properties that are not relevant for our generic discussion.

We begin by writing an integral equation for $A(\vec p)$, whose iterative solution 
corresponds to the sum of the ladder diagrams in which the incoming particles repeadly 
exchange vector bosons before annihilating.
Then we take the non-relativistic limit finding an in-homogeneous Schroedinger equation 
relating the Fourier transform of  $A(\vec p)$ to the Fourier transform of $A_0(\vec p)$.
The solution of this equation gives:
\be
A(\vec p)=\int d\vec r\phi^*_{\vec p}(\vec r)
\int {d\vec q\ov (2\pi)^3} e^{i\vec q\cdot\vec r}A_0(\vec q)
\ee
 where $\phi_{\vec p}$, normalized as $\int d\vec r\phi^*_{\vec p}(\vec r)\phi_{\vec k}(\vec r)=(2\pi)^3\delta(\vec p-\vec k)$,
solves the Schroedinger equation 
\be
(-{1\ov 2m_r}\p^2-{\alpha e^{-\mu r}\ov r}-{p^2\ov 2m_r})\phi_{\vec p}(\vec r)=0
\label{mainS}
\ee
$m_r=m/2$ is the reduced mass, $\mu$ the vector potential mass and $\alpha=g^2/(4\pi)$
its strength. 

This formula for the complete amplitude $A$ can be numerically evaluated 
once the dependence in $\vec p$ of $A_0(\vec p)$ is known.

Furthermore, we consider the case in which the annihilation 
occurs at a definite partial wave $l$ and therefore assuming that for $p$ small 
$A_0\sim p^l$, i.e. the typical dependence on $p$ of the $l$ partial wave.

We define the 
enhancement $enh_l$ as the factor that multiplies the $bare$ cross-section $\sigma_{0,l}$
to give the complete one $\sigma_l$ (for the rate it would be the same):
\be
\sigma_l=enh_l\cdot \sigma_{0,l}
\ee
We find that $enh_l$ depends, beside $l$, on the two dimensionless parameters:
\be
a:= {\alpha\ov v},~~~ b:={\mu\ov m_r v}
\ee
where $m_r=m/2$ and $v$ is the relative CM velocity.

Analytical computations are possible
in the case in which $\mu$ is negligible.
We find for the generic $l$ partial wave :
\be
enh_l=\prod_{s=1}^l(s^2+a^2) e^{\pi a}{\pi a \ov sinh(\pi a)~ l!^2}
\ee
In particular,in the limit $a>>1$, for the S wave we find the standard 
Sommerfeld enhancement $2\pi{\alpha\ov v}$, \cite{LL},\cite{Hisa},\cite{ Stru},\cite{Mar}, \cite{ArkHam},\cite{Lat}, 
whereas  for the P ($l=1$) wave we find the enhancement $2\pi({\alpha\ov v})^3$.
\vskip0.2cm
In the general case of a Yukawa potential, we derive a formula expressing $enh_l$ in terms of the asymptotic 
behavior of the partial wave solution of the homogeneous Schroedinger equation, and we indicate a very easily implementable 
algorithm to evaluate it numerically, which does not require any computational skill.


The case of the S $l=0$ wave has been discussed at length in the literature, and it has been found 
a resonant pattern see in particular refs. \cite{Lat} and \cite{Rob}.

Here we present some numerical results for the P $l=1$ wave, which also shows resonances. In this case the enhancement can be 
of several orders of magnitude also away from the resonance, and much more on it.

\vskip0.5cm

\section{Derivation of the results}

 \setcounter{equation}{0}

\subsection{The equation for the  amplitude} 

Define $A(p,p';P_0)$ to be the amplitude for the annihilation process of two 
$\chi$ particles

$$
\chi(p_1)+\bar\chi(p_2) \to a(p'_1)+\bar a(p'_2)
$$
The final state $a,\bar a$ can be any (quantum-number compatible) two particle state of the standard model.

Define $P={p_1+p_2\ov 2}={p'_1+p'_2\ov 2}$ , $p={p_1-p_2\ov 2},~p'={p'_1-p'_2\ov 2}$.
In the CM $P_0=\sqrt{p^2+m^2},~\vec P=0,~p_0=0$.

\vskip0.2cm

$A(p,p';P_0)$ is the complete amplitude, including the Sommerfeld effect, and we call
$A_0(p,p';P_0)$ the "bare" amplitude, that is neglecting the Sommerfeld effect. The variable $p'$ does not
play any role in the following, we continue to indicate it just for completeness.

Here we treat  the initial particles $ \chi$ having a mass $m$
as Dirac particles of opposite charge, to be definite, and imagine that before annihilating 
they attract each other by the exchange of a vector boson called VB (the exchange of a scalar would give
the same non-relativistic potential).
We take the general case of the VB with mass $\mu<<m$. In the results one can put $\mu=0$.
In the non-relativistic limit, any kind of $\chi$ particle, Dirac, Majorana or even scalar, would give the same result. 
As said, in the case of non-abelian interaction, we assume that it has been diagonalized. 

\vskip0.2cm

Here we follow closely the Chapter 10 of the book "Quantum Field Theory" by
C.Itzykson and J.B.Zuber \cite{IZ}.

Since there are two fermions in the initial state we distinguish the Dirac matrices
acting on the two particles by a suffix, say $\gamma_1,\gamma_2$ act on particle
$1$ and $2$ respectively.

The VB exhange between the two incoming particles gives a factor
(vertex-propagator-vertex) 
$$
{\gamma^0_1\gamma^0_2-\vec\gamma_1\cdot\vec\gamma_2\ov  k^2-\mu^2}=
-{\gamma^0_1\gamma^0_2\ov \vec k^2+\mu^2}+
({\gamma^0_1\gamma^0_2k_0^2\ov (k^2-\mu^2)(\vec k^2+\mu^2)}
-{\vec\gamma_1\cdot\vec\gamma_2\ov k^2-\mu^2})
$$

where $k=p-p'$. The first term in the r.h.s. is the instantaneous "Coulomb like" interaction and it will treated non-perturbatively, whereas the second term, containing the retarded and magnetic effects 
can eventually be included as a perturbation \cite{IZ} and it 
will be ignored in the following discussion. 

Also, we will include only the ladder diagrams, that is the iteration of the one VB exchange \cite{IZ}. 

The reason for it is that the complete amplitude $A(p,p';P_0)$ 
will be determined, in the relevant non-relativistic approximation, by the non-perturbative
solution of a Schroedinger equation with a potential to be determined by the field theory diagrams.
A usual strategy for solving the Schroedinger equation is, whenever possible, to split the potential
into a dominant part which can be exactly solved, plus higher order terms giving corrections to be
computed, if necessary, by perturbation theory, like it is done for the fine and
hyperfine corrections to the energy levels of the hydrogen atom.

In our case the general diagrammatic expansion of the amplitude can be seen as 
the iteration of two-particle irreducible sub-diagrams and the non-relativistic potential corresponds to 
the sum of those two-particle irreducible diagrams.
The dominant term of the  potential, for which we will provide the exact solution,
comes from the lowest order two-particle irreducible diagram, that is the one VB exchange.

In the general expansion it appears also the iteration of other two-particle irreducible
sub-diagrams, including radiative corrections and sub-diagrams in which the VB lines cross;  as said,
they correspond to additional contributions to the interaction, which are of higher order in the coupling constant, i.e. higher than the one-VB potential, and their effect , if necessary, can be computed by perturbation theory.
This is similar to the treatment of the positronium in ref.\cite{IZ}, in which the  Bethe-Salpeter equation in the ladder approximation 
provides the leading non-perturbative solution, with perturbative corrections coming from higher order diagrams.
 
Further, we also note that  the crossed diagrams are less singular for $k^2\sim O(\mu^2)$  than the one-VB exchange;
since the range of the interaction is determined by the strength of the nearby singularity, they will correspond 
to an interaction which is less effective in attracting the incoming particles at large distances, beside giving
 a small correction to the exact solution solution of the one-VB potential, because
they represent higher orders in the interaction.

\vskip0.2cm

That being said, $A$ satisfies the following integral equation ($\hat p\equiv\gamma^\mu p_\mu$):
\bea
A(\tilde p,p';P_0 ) = A_0(\tilde p,p';P_0) - \\ \nonumber
-   i g^2\int {d^3 q d q_0\ov (2\pi)^4} {\gamma^0_1\gamma^0_2 \ov (\vec{\tilde p}-\vec q)^2 +\mu^2}
{(\hat P+\hat q+m)_1\ov (P+q)^2-m^2+i\epsilon}{(\hat P-\hat q+m)_2\ov (P-q)^2-m^2+i\epsilon}
A(q,p';P_0)
\label{ladder}
\eea
Its iterative solution corresponds to the sum of the ladder diagrams in which the incoming particles repeadly 
exchange VB's before annihilating.
After solving the equation one has to put $\tilde p= p$, that is $\tilde p_0=p_0=0,~\vec{\tilde p}=\vec p$.

We evaluate the integration over $q_0$ by closing the contour in the complex plane
disregarding the possible singularities of $A$ which give sub-leading terms in the non-relativistic limit
(see Appendix A). In the following $v:=|\vec v|$  denotes the modulus of a 3-vector.

The poles in the lower plane are located at :
$q_0=\omega-P_0-i\epsilon, ~~~ q_0=\omega+P_0-i\epsilon$
with $\omega=\sqrt{q^2+m^2}$.
We write \cite{IZ}
\bea
{(\hat P+\hat q+m)_1\ov (P+q)^2-m^2+i\epsilon}{(\hat P-\hat q+m)_2\ov (P-q)^2-m^2+i\epsilon} = 
~~~~~~~~~~~~~~~~~~~~~~~~~~~~~~~~~~~~~~~~~~~~~~~~~\\ \nonumber
({\Lambda^{+}_1(\vec q)\ov q_0+ P_0-\omega+i\epsilon}+
{\Lambda^{-}_1(\vec q)\ov q_0+ P_0+\omega-i\epsilon})\gamma^0_1 
({- \Lambda^{-}_2(- \vec q)\ov q_0- P_0-\omega+i\epsilon}+
{-\Lambda^{+}_2(- \vec q)\ov q_0- P_0+\omega-i\epsilon})\gamma^0_2
\eea

where $\Lambda^{\pm}={\omega \pm H\ov 2\omega}$ and 
$H=\vec\alpha\cdot\vec q+\beta m$ with $\beta=\gamma^0,~\alpha=\beta\gamma$.

\vskip0.2cm

The residue at $q_0=\omega-P_0$ is
\be
\Lambda_1^{+}(\vec q)
({\Lambda_2^{-}(-\vec q)\ov 2P_0}-{\Lambda_2^{+}(-\vec q)\ov 2(\omega-P_0)})
\gamma^0_1\gamma^0_2
~~\rightarrow~~ -{\Lambda_1^{+}(\vec q)\Lambda_2^{+}(-\vec q)\ov 2(\omega -P_0)}\gamma^0_1\gamma^0_2
\ee

here we take the leading term in the non-relativistic limit, that is the term containing at the denominator 
$\omega -P_0=\sqrt{q^2+m^2}-\sqrt{p^2+m^2}$ (small for $m\to\infty$).

The residue at $q_0=\omega+P_0$ is
\be
({\Lambda_1^{+}(\vec q)\ov  2P_0}+{\Lambda_1^{-}(\vec q)\ov 2(\omega+P_0)})(-\Lambda_2^{-}(-\vec q))\gamma^0_1\gamma^0_2
\ee

here there is no small denominator, and we neglect it as a sub-leading term.

In the non-relativistic limit $\gamma^0\sim1$ and  $\Lambda_1^{+}(\vec q)\Lambda_2^{+}(-\vec q)\sim1$ (leading term for $m\to\infty$).
Futher, in the nonrelativistic limit, $A(\omega-P_0,\vec q,p';P_0)\sim  A(0,\vec q,p';P_0)$.
Therefore we can consistently put directly $\tilde p_0=0$ and consider 
the equation for $A(\vec{\tilde p},p';P_0)\equiv A(0,\vec{\tilde p},p';P_0)$.

In conclusion, in the non-relativistic approximation we get
\be
A(\vec{\tilde p},p';P_0)=A_0(\vec{\tilde p},p';P_0)+{g^2\ov (2\pi)^3}
\int {d^3 q   \ov (\vec{\tilde p}-\vec q)^2+\mu^2 }
{A(\vec q,p';P_0)\ov  2(\omega -P_0)}
\ee

 In the denominator $\omega -P_0$ we take the leading term for $m$ large compared to the three-momentum:
 \be
 {1\ov 2(\omega-P_0)}\to {1\ov { q^2\ov 2m_r}-{\cal E}}
 \ee
 where ${\cal E}=2(P_0^2-m^2)/(2m)= p^2/(2m_r)$  is the total non-relativistic energy
 ($m_r=m/2$ is the reduced mass).

The final step is a further redefinition:
\be
A(\vec{\tilde p},p';P_0)=({{\tilde p}^2\ov 2m_r}-{\cal E})\tilde\Psi_{\cal E}(\vec{\tilde p},p')
\ee
getting the non-relativistic equation
\be
({{\tilde p}^2\ov 2m_r}-{\cal E})\tilde\Psi_{\cal E}(\vec{\tilde p},p')= A_0(\vec{\tilde p},p';P_0)+{g^2 \ov (2\pi)^3}\int {d^3 q   \ov (\vec{\tilde p}-\vec q)^2+\mu^2 }\tilde\Psi_{\cal E}(\vec q,p')
\ee

Since  ${1\ov k^2+\mu^2}={1\ov 4\pi}\int d\vec r~{e^{-i\vec k\cdot\vec r-\mu r}\ov r}~$, 
defining  $\alpha:=g^2/(4\pi)$ and
\be
 \Psi_{\cal E}(\vec r):=\int d\vec{\tilde p} e^{i\vec{\tilde p}\cdot\vec r}\tilde\Psi_{\cal E}(\vec{\tilde p},p'),
~~~U_0(\vec r):=\int d\vec{\tilde p} e^{i\vec{\tilde p}\cdot\vec r}\tilde A_0(\vec{\tilde p},p',P_0)
\ee

we get  the in-homogenous equation:
\be
(-{1\ov 2m_r}\p^2-{\alpha e^{-\mu r}\ov r}-{\cal E})\Psi_{\cal E}(\vec r)=U_0(\vec r)
\label{general}
\ee

\subsection{Solution of the equation}

Eq.(\ref{general}) is formally solved by 
\be
\Psi_{\cal E}(\vec r)=\int d\vec r' 
\int {d\vec k\ov (2\pi)^3} 
{\phi_{\vec k}(\vec r)\phi^*_{\vec k}(\vec r')\ov {k^2\ov 2m_r}-{\cal E}-i\ep} U_0(\vec r')
\ee
(we choose the Feynmann contour prescription: $\vec k^2\to\vec  k^2-i\ep$, for our purpose  $+i\ep$ would give the same) 
where $\phi_{\vec  k}(\vec r)$ are a complete set of solution of
\be
(-{1\ov 2m_r}\p^2-{\alpha e^{-\mu r}\ov r}-{k^2\ov 2m_r})\phi_{\vec k}(\vec r)=0
\label{sch}
\ee

normalized such that the completeness relation is
$\int {d\vec k \ov (2\pi)^3}\phi_{\vec k}(\vec r)\phi^*_{\vec k}(\vec r')=\delta(\vec r-\vec r')$.

The final step is to reconstruct
 $A_{on-shell}(\vec p,p';P_0)=\lim_{\vec{\tilde p}\to\vec p} A(\vec{\tilde p},p';P_0)$ on the mass-shell,
that is for  ${{\tilde p}^2 \ov 2m_r}\to {\cal E}={p^2\ov 2m_r}$ i.e.  ${\tilde p}^2\to P_0^2-m^2=p^2$.
We get
\bea
A_{on-shell}(\vec p,p';P_0)=\lim_{\vec{\tilde p}\to\vec p}~({{\tilde p}^2 \ov 2m_r}-{\cal E})
\tilde\Psi_{\cal E}(\vec{\tilde p},p') \\ \nonumber
={1\ov (2\pi)^3}\lim_{\vec{\tilde p}\to\vec p}
~( {\tilde p}^2- p^2)\int  {d\vec k\ov (2\pi)^3}
{[\int d\vec r e^{-i\vec{\tilde p}\cdot \vec r} \phi_{\vec k}(\vec r) ]
[\int d\vec r' \phi^*_{\vec k}(\vec r') U_0(\vec r')] \ov   k^2- p^2-i\ep} 
\eea

We use the formula (see Appendix B)

\be
{1\ov (2\pi)^3}\lim_{\vec{\tilde p}-\vec p}~( {\tilde p}^2-p^2)
{\int d\vec r e^{-i\vec{\tilde p}\cdot \vec r} \phi_{\vec k}(\vec r)\ov k^2-p^2-i\ep} =
\delta(\vec p-\vec k)
\label{final}
\ee

Then the scattering amplitude $A$ on shell  including the Sommerfeld effect turns out to  be
\be
A_{on-shell}(\vec p,p';P_0)=\int d\vec r\phi^*_{\vec p}(\vec r)
\int {d\vec q\ov (2\pi)^3} e^{i\vec q\cdot\vec r}A_0(\vec q,p';P_0)
\label{main}
\ee
This result can be rewritten in a trasparent way as
\be
A_{on-shell}(\vec p,p';P_0)=\int d\vec q \tilde\phi^*_{\vec p}(\vec q)<\vec q,-\vec q|M|p'>
\ee
where $<\vec q,-\vec q|M|p'>=A_0(\vec q,p')$ is the matrix element of the annihilation reaction, and $ \tilde\phi_{\vec p}(\vec q)$ 
is the momentum-space wave function of the incoming pair, which takes into account the mutual interaction, normalized such that in absence of
interaction $A=A_0$. (A formula of that kind is heuristically presented in the Chapter 5  of the Peskin and Schroeder book \cite{Pesk}. With our Feynmann graph 
convention the initial state appears to the left rather than to the right).

Let us see how the above general formula works in two particular cases.

In the case in which $A_0$ is S-wave dominated and then it is a constant, we have:
\be
A_0(\vec p,p';P_0)=a_0 \rightarrow \int {d\vec q\ov (2\pi)^3} e^{i\vec q\cdot\vec r}A_0(\vec q,p';P_0)=\delta(\vec r)a_0
\ee
Therefore by eq.(\ref{main})
\be
A_{on-shell}(\vec p,p';P_0)=\phi^*_{\vec p}(0)a_0
\ee

In the case $A_0(\vec q,p';P_0)$ is P-wave dominated then it is linear in $p$
\be
A_0(\vec q,p';P_0)=\vec p\cdot\vec p' a_1  \rightarrow
 \int {d\vec q\ov (2\pi)^3} e^{i\vec q\cdot\vec r}A_0(\vec q,p';P_0)= -i   \vec\p\delta(\vec r)\cdot\vec p' a_1
\ee
therefore by eq.(\ref{main}) 
\be
A_{on-shell}(\vec p,p';P_0)=i \vec\p\phi^*_{\vec p}(0)\cdot\vec p' a_1
\ee

To make explicit computations, one has to find $\phi$ around $r=0$. In order to do that,  it is convenient first 
to specialize the general formula (\ref{main}) for the case of a definite partial wave $l$.

\section{The enhancement formula for generic l-waves}

The wave function $\phi_{\vec p}(\vec r)$ can be decomposed in partial waves \cite{LL}:
\be
\phi_{\vec p}(\vec r)={(2\pi)^{3/2}\ov 4\pi p}\sum_l i^l(2l+1)e^{i\delta_l}R_{p,l}(r)P_l(\hat p\cdot\hat r)
\label{partwexp}
\ee
($\hat p,\hat r$ are unit vectors in the direction of $\vec p,\vec r$). 

$R_{p,l}(r)$ is the solution of the partial wave Schroedinger equation:
\be
-{1\ov 2m_r}({d^2 R_{p,l}\ov dr^2}+{2\ov r}{dR_{p,l}\ov dr}-{l(l+1)R_{p,l}\ov r^2})-
({p^2\ov 2m_r}+{\alpha e^{-\mu r}\ov r})R_{p,l}=0
\label{eqpartwave}
\ee
normalized such that 
\be
\int_0^\infty r^2 dr R_{q,l}(r)R_{p,l}(r)=\delta (p-q)
\label{normpartwave}
\ee
and the completeness is 
\be
\int_0^\infty dp R_{p,l}(r) R_{p,l}(r')={1\ov r^2}\delta(r-r')
\label{complete}
\ee
(This can be checked to be consistent with the  normalization 
$\int d\vec r\phi^*_{\vec p}(\vec r)\phi_{\vec k}(\vec r)=(2\pi)^3\delta(\vec p-\vec k)$
and with the completeness
$\int  d^3\vec k \phi_{\vec k}(\vec r)\bar\phi_{\vec k}(\vec r')=(2\pi)^3\delta^3(\vec r-\vec r')$).
We have taken the convention that $R_{p,l}(r)$ is real, which we can always do.

For the free case $\phi^0_{\vec p}(\vec r)=e^{i\vec p\vec r}$ and $ R^0_{p,l}(r)=\lim_{\alpha\to 0}R_{p,l}(r)$. 

We use the following identities \cite{LL}:
\be
\int d\Omega_k P_l(\hat k\cdot\hat r)P_{l'}(\hat k\cdot\hat r')={4\pi\ov 2l+1}\delta_{ll'}P_l(\hat r\cdot\hat r')
 ~~~~~~~~~~~~~~~~~~~~~~~~~ \label{partid}
 \ee
 \be
\sum_l(2l+1)P_l(\hat r\cdot\hat r')=2\delta(1-\hat r\cdot\hat r') ~~~
\delta^3(\vec r-\vec r')={1\ov r^2}\delta(r-r'){1\ov 2\pi}\delta(1-\hat r\cdot\hat r') 
\ee

Take for the $l$ wave :  \\
$~~A_l(\vec p,\vec p')=A_l(p,p') P_l(\hat p\cdot\hat p')~~$ and similarly 
$~~A_{0,l}(\vec q,\vec p')=A_{0,l}(q,p') P_l(\hat q\cdot\hat p')$. 

\vskip0.2cm

Putting the expansion eq.(\ref{partwexp}) in the main formula eq.(\ref{main})
 and using eq.(\ref{partid}) to do the angular integration, we get the partial wave
 version of eq.(\ref{main}) 
 \be
 A_l(p,p')={1\ov p} \int_0^\infty r^2 dr  R_{p,l}(r)\int_0^\infty qdq R^0_{q,l}(r)A_{0,l}(q,p')
 \ee
  
 The standard dependence on small $q$ is $A_{0,l}(q,p')=q^l a_{0,l}(p')$. Since \cite{LL}
 \be
 ({d\ov dr})^lR^0_{q,l}(r)|_{r=0}= \sqrt{2\ov \pi}{l!\ov 1\cdot 3\cdots (2l+1)} q^{l+1}
 \ee
 using the completeness eq.(\ref{complete}) for the free $R^0_{q,l}$ we get
 \be
 \int q dq R^0_{q,l}(r) q^l ~a_{0,l}(p')
 =(-)^l\sqrt{\pi\ov 2}{1\cdot 3\cdots (2l+1)\ov l!}{1\ov r^2}\delta^l(r)~a_{0,l}(p')
 \ee
 
 In conclusion we get
 \bea
 A_l(p,p') &=& (-)^l\sqrt{\pi\ov 2}{1\cdot 3\cdots (2l+1)\ov l!}{1\ov p}
 \int r^2 dr R_{p,l}(r){1\ov r^2}\delta^l(r)~a_{0,l}(p') \\ 
 &=&  \sqrt{\pi\ov 2}{1\cdot 3\cdots (2l+1)\ov l!}{1\ov p}~({d\ov dr})^l R_{p,l}(r)|_{r=0}~a_{0,l}(p')
 \label{enhpartwave}
 \eea
 
\section{The enhancement for the case of a Coulomb potential} 
 
 The Coulomb partial wave is \cite{LL}, with our notation $a:=\alpha/v$ ($v$ relative CM velocity), 
\be
R^c_{p,l}(r)=p\sqrt{{2\ov\pi} \prod_{s=1}^l (s^2+a^2)}e^{\pi a\ov 2} \Gamma (1-i a)
~{(2pr)^{l}e^{-ipr}\ov (2l+1)!}F({i a}+l+1,2l+2,2ipr)
\label{coulpart}
\ee

It is easy to evaluate $({d\ov dr})^l  R^c_{p,l}(r)|_{r=0}$ by using the explicit expression 
eq.(\ref{coulpart}) since for $r\to 0$ only the $l$-time derivative of the factor $(2pr)^l$  contributes
and $F(a,b,2ipr)|_{r=0}=1$.
 
 We get 
\footnote{checking the normalization: for $\alpha\to 0$ one gets $ A_l(p,p')=p^la_{0,l}(p')$} :
 \be
 A_l(p,p')=\sqrt{\prod_{s=1}^l (s^2+a^2)}e^{\pi a\ov 2} \Gamma (1-i a) {1\cdot 3\cdots (2l+1)\ov (2l+1)!}2^l~p^la_{0,l}(p')=
 {\sqrt{\prod_{s=1}^l (s^2+a^2)}e^{\pi a\ov 2} \Gamma (1-i a)\ov l!}~p^la_{0,l}(p')
 \ee
 
 Since $p^la_{0,l}(p')=A_{0,l}(p,p')$, and  $|\Gamma(1-i/p_c)|^2={\pi a\ov sinh(\pi a)}$
 by taking the square modulus  we get the enhancement formula 
 $\sigma_l=enh_l \cdot \sigma_{0,l}$ 
 where 
 \be
 enh_l=\prod_{s=1}^l(s^2+a^2) e^{\pi a}{\pi a\ov sinh(\pi a) l!^2}
 \label{enhcoul}
 \ee 
  For large $a=\alpha/v$, $enh_l={2\pi \ov l!^2}({\alpha\ov v})^{2l+1}$.

\section{The enhancement for the case of a Yukawa potential}

Here we take eq.(\ref{enhpartwave}) and insert for $R_{p,l}(r)$ the solution of eq.(\ref{eqpartwave}), normalized as 
eq.(\ref{normpartwave}).

We know from textbooks (see for instance \cite{LL}) that this normalization corresponds to
the asymptotic behavior
\be
{R_{p,l}(r)}_{r\to\infty} \to \sqrt{2\ov\pi}{\sin(pr-{l\pi\ov 2}+\delta_l)\ov r}
\label{normR}
\ee

Let us define $x=pr$ and put $R_{p,l}(r)=N px^l\Phi_l(x)$; the equation for $\Phi_l$ is:
\be
\Phi_l''+{2(l+1)\ov x}\Phi_l'+({2 a e^{-bx}\ov x}+1)\Phi_l=0
\label{eqphi}
\ee
where again $a:= \alpha /v$, $b:= \mu /(m_r v)$ and $v=p/m_r$ is the relative velocity. 

Suppose we solve this equation with the initial conditions 
\be
\Phi_l(0)=1 ~~~~~ \Phi_l'(0)=-a/(l+1)
\label{init}
\ee
(the condition for $\Phi_l'(0)$ is dictated by the equation
for a regular solution). Then the asymptotic behavior will be
\be
{x^{l+1}\Phi_l(x)}_{x\to\infty}\to C\sin(x-{l\pi\ov 2}+\delta_l)
\label{asy}
\ee
In order to agree with the normalization of eq.(\ref{normR}) we have to put 
$N=\sqrt{2\ov\pi}{1\ov C}$. Substituting in eq.(\ref{enhpartwave}) we get
\be
A_l(p,p')={1\cdot 3\cdots (2l+1)\ov C} p^la_{0,l}={1\cdot 3\cdots (2l+1)\ov C} A_{0,l}(p,p')
\ee
In conclusion, by defining the Sommerfeld enhancement $enh_l$ for the $l$ partial wave cross-section (or equivalently for the rate) as
\be
\sigma_l=enh_l\cdot\sigma_{0,l}
\ee
 we get
\be
enh_l=({1\cdot 3\cdots (2l+1)\ov C})^2
\label{result}
\ee
where $C$ is obtained by looking at the asymptotic behavior eq.(\ref{asy}) of the solution of
eq.(\ref{eqphi}) with the initial conditions eq.(\ref{init}). 
$enh_l$ depends on the two parameters $a$ and $b$. It is not necessary to determine $\delta_l$.

Another equivalent  strategy is to put $R_{p,l}(r)=N p\varphi_l(x)/x$; the equation for $\varphi_l$ is:
\be
\varphi_l''+(1+{2 a\ov x}e^{-bx}-{l(l+1)\ov x^2})\varphi_l=0
\label{eqb}
\ee
If one solves this equation with the initial conditions corresponding to 
\be
{\varphi_l(x)}_{x\to 0}\to x^{l+1}
\label{initb}
\ee 
then the 
asymptotic behavior will be
\be
{\varphi_l(x)}_{x\to\infty}\to C\sin(x-{l\pi\ov 2}+\delta_l)
\label{asyb}
\ee
with the same $C$ of eq.(\ref{asy}) giving the enhancement as in eq.(\ref{result}).

\subsection{Computations for $l=1$}

In principle it is easy to get $C$: for instance one can use the NDSolve instruction of 
Mathematica to get the numerical solution of eq.(\ref{eqphi}) with the initial 
conditions eq.(\ref{init}), or equivalently of eq.(\ref{eqb})  with initial conditions eq.(\ref{initb}).
In order to find $C$ one takes $F_l(x)\equiv x^{l+1}\Phi_l(x)$ or else $F(x)\equiv \varphi(x)$,
and  one
plots $F_l(x)^2+F_l(x-\pi/2)^2$ for large $x$: when this is constant it is equal to $C^2$. 
We follow the strategy of eqs.(\ref{eqb},\ref{initb}), which provides more clean numerical results.

It is expected that this procedure should work less well for $b$ very low and $a$ very large because
in this case the asymptotia is reached for very large $x$ and the numerical solution
accumulates errors. However, for $b=0$ and for any $a$ we already have  the exact result derived analytically,
eq.(\ref{enhcoul}).

In practice, this works well for $l=0$, see for instance Fig. 1.

\begin{figure}[ht]\label{fig.1}
\centerline{\epsfig{file=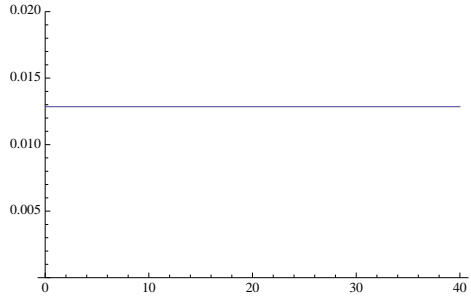,width=.4\textwidth}}
\caption{$F_0(x)^2+F_0(x-\pi/2)^2$ for $\alpha=1/100,\mu=1Gev,m=10^3Gev,v=2\times 10^5$}
\end{figure}

For $l=1$ the quantity $F_l(x)^2+F_l(x-\pi/2)^2$ may sometimes continue to show
decreasing oscillations: here it is convenient to take into account the sub-leading term in the asymptotic
expansion (\ref{asyb}), which we know to be the free wave function up to the phase shift $\delta_l$.
Therefore, the improved version of (\ref{asyb}) for $l=1$ is:
\be
{\varphi_1(x)}_{x\to\infty}\to C\cdot \big(\sin(x-{l\pi\ov 2}+\delta_l)+{\cos(x-{l\pi\ov 2}+\delta_l)\ov x}\big)
\ee
To get $C^2$, one defines, in three steps, with $F_1(x):= \varphi_1(x)$,  
\bea
k(x) &:=& {(\pi^2-16 x^2)^2\ov(8\pi(\pi^2-16x^2))}(F_1(x+\pi/4)^2+F_1(x-\pi/4)^2), \nonumber \\ 
h(x) &:=& (\pi^3-4\pi x^2)(k(x+\pi/4)+k(x-\pi/4), \nonumber \\
j(x) &:=&-8{h(x+\pi/4)+h(x-\pi/4)\ov  (\pi^2-16x^2)(3\pi^2+16(1+x^2))}. \nonumber
\eea
For $x$ large (say $x>30$), $j(x)$ quickly converges to a constant  which  equals $C^2$.  
(For $l>1$ the free wave function is more complicated and one should do more steps).
It must be said that the results of the improved procedure differ little
from what could be obtained simply by finding by eye the average of the 
oscillations of  $F_l(x)^2+F_l(x-\pi/2)^2$, see Fig. 2.

\begin{figure}[ht]\label{fig.2,a,b}
\centerline{\epsfig{file=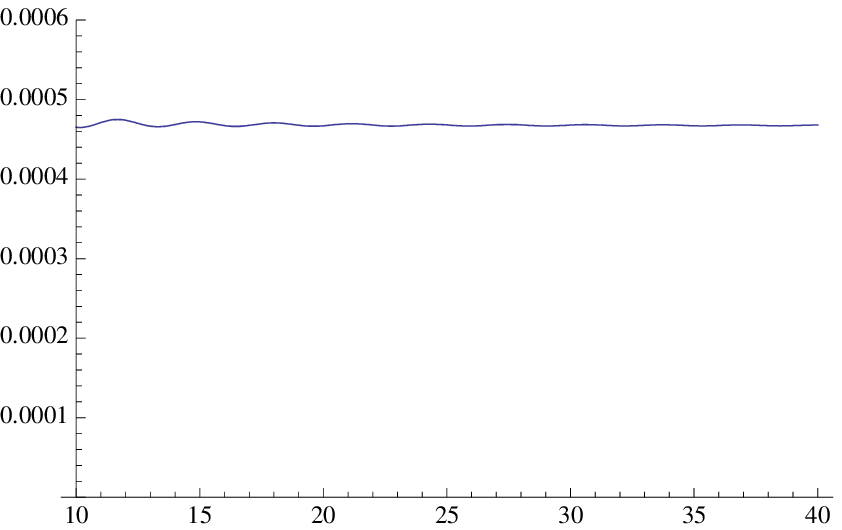,width=.4\textwidth}\ \epsfig{file=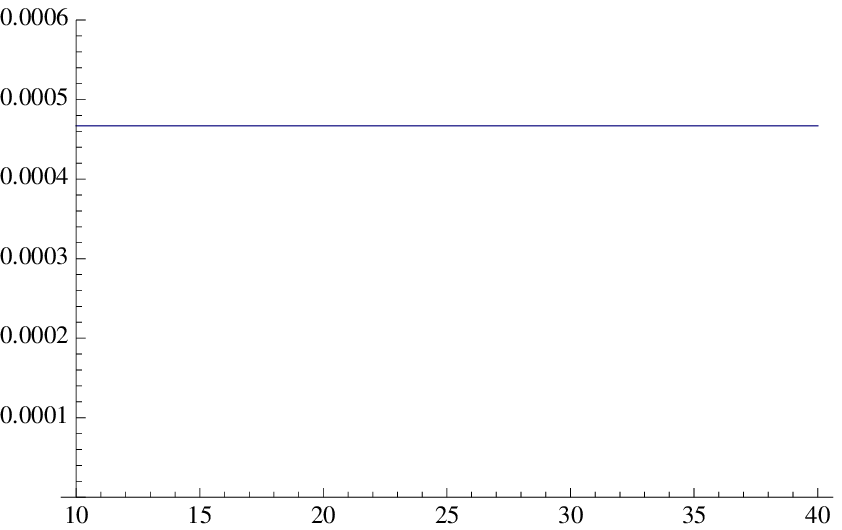,width=.4\textwidth}}
\caption{ $F_1(x)^2+F_1(x-\pi/2)^2$ (left) and $j(x)$ (right) for the same parameters of Fig.1}
\end{figure}

\vskip1cm

The case of the S $l=0$ wave has been discussed at length in the literature, and it has been found 
a resonant pattern, see in particular refs. \cite{Lat} and \cite{Rob}. We have verified that we get the same results.

Here we present some numerical result for the P $l=1$ wave, which also shows a resonant pattern.

In Fig.3 we show the enhancement eq.(\ref{result}) for $l=1$, taking the values of the parameters 
used in the numerical evaluations for the S wave in ref. \cite{Lat} and also reported in ref. \cite{Rob}, 
that is $\alpha=1/30,~\mu=90Gev$, as a function of $m$ (expressed in $Gev$)
for $v_{single~particle}=10^{-3},10^{-4},10^{-5}$.

In Fig.4 we show the enhancement eq.(\ref{result}) for $l=1$, taking the values of the parameters 
used in the numerical evaluations for the S wave in ref. \cite{Rob} for the range of ref. \cite{ArkHam}, that is 
$\alpha=1/100,~\mu=1Gev$, as a function of $m$ (expressed in $Gev$)
for $v_{single~particle}=10^{-3},10^{-4},10^{-5}$.


\begin{figure}[ht]\label{fig.3}
\centerline{\epsfig{file=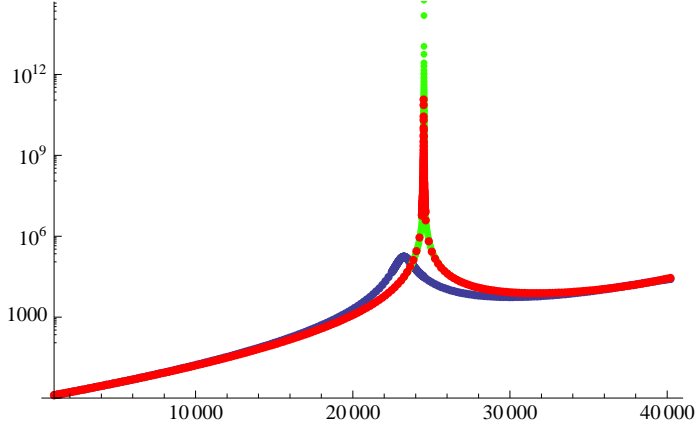,width=.6\textwidth}}
\caption{ $enh_1$ as a function of $m(Gev)$ for $v_{single-particle}=10^{-3}$ blue, $10^{-4}$ red ,
$10^{-5}$ green. Here $\alpha=1/30$, $\mu=90 Gev$.}
\end{figure}

\newpage

\begin{figure}[ht]\label{fig.4}
\centerline{\epsfig{file=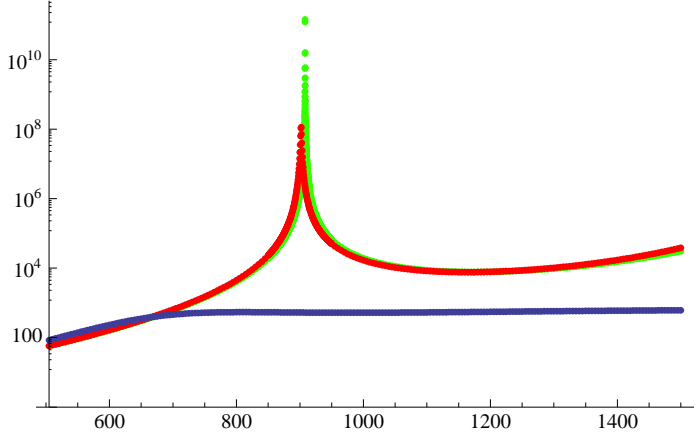,width=.6\textwidth}}
\caption{ $enh_1$ as a function of $m(Gev)$ for $v_{single-particle}=10^{-3}$ blue, $10^{-4}$ red ,
$10^{-5}$ green. Here $\alpha=1/100$, $\mu=1 Gev$.}
\end{figure}


\appendix

\section{The singularities in $q_0$ of the integrand at the r.h.s of eq.(\ref{ladder}) }
 \setcounter{equation}{0}

Consider the iteration of eq.(2.1). At the first order we put $A_0$ in the place of $A$
in the r.h.s. Since $p'$ is not touched in the integral equation, we can directly 
take its mass-shell value: \\
$p'_0=\omega(p),\vec p'=\omega(p)\vec n$ ,
$\omega(p)=\sqrt{p^2+m^2}$ (treating
the final standard-model particles as massless). 

The singularities of $A_0$ come from the denominator of the propagator exchanged
between the vertex $\chi (q_1) a(p'_1)$ and the vertex $\bar\chi(q_2)\bar a( p'_2)$ 
(we take a massive propagator with a mass $\hat m$ of the order  of -maybe equal to-
 $m$):
\be
(q-p')^2-\hat m^2+i\ep=[q_0-(u(q,p)+\omega(p))+i\ep][q_0+(u(q,p)-\omega(p))-i\ep]
\ee
where $u(p,q)= \sqrt{\omega(p)^2+\hat\omega(q)^2-2\vec q\cdot\vec n\omega(p)}$
and $\hat\omega(q)=\sqrt{q^2+\hat m^2}$.

.
 
 Therefore the integration over $q_0$ in eq.(\ref{ladder}) is (remember $P_0=\omega(p),\vec P=0$)
 \bea
 \int dq_0 N(q_0,\vec q,\vec p)\times {1\ov [q_0-(\omega(q)-\omega(p))+i\ep][q_0+(\omega(q)+\omega(p))-i\ep]} \times \\ \nonumber
 {1\ov [q_0-(\omega(q)+\omega(p))+i\ep][q_0+(\omega(q)-\omega(p))-i\ep]} \times \\ \nonumber
 {1\ov [q_0-(u(q,p)+\omega(p))+i\ep][q_0+(u(q,p)-\omega(p))-i\ep]}
 \eea
 
where $N$ is some numerator, polynomial in $q_0$.
We do the integration on $q_0$ in eq.(A.2) by closing the contour in the lower half-plane.

The relevant poles are 
\bea
(1)~q_0 &=& \omega(q)-\omega(p) \\ \nonumber
 (2)~q_0 &=& \omega(q)+\omega(p) \\ \nonumber
(3)~q_0 &=& \sqrt{\omega(p)^2+\hat\omega(q)^2-2\vec q\cdot\vec n\omega(p)}+\omega(p)
\eea

By computing the residue of the countour integration, it is seen that only the residue of the pole $(1)$ 
contains in the denominator the small factor $\omega(q)-\omega(p)\to 0$
(for ~$m\to\infty$),
providing the leading contribution in the non-relativistic limit. Therefore, in this limit
it is tantamount to do the contour integration disregarding the singularities of $A_0$, 
that is of $A$ at the first order iteration.

This remains true for the next iterations since the instantaneous part of the VB propagator
(that is the step which is making the ladder) is $q_0$ independent.

\section{Derivation of the formula eq.(\ref{final})}
 \setcounter{equation}{0}

Eq.(\ref{sch}) can be recast as an integral equation (remembering our choice $k^2-i\ep$) 
\be
\phi_{\vec k}(\vec r)=e^{i\vec k\cdot\vec r}+ \eta_{\vec k}(\vec r)
~~~where~~~
\eta_{\vec k}(\vec r)=
{1\ov -\p^2-k^2+i\ep}\cdot {2m_r\alpha e^{-\mu r}\ov r} \phi_{\vec k}(\vec r)
\ee

Therefore
\bea
{1\ov (2\pi)^3}\lim_{\vec{\tilde p}-\vec p}( {\tilde p}^2-p^2)
{\int d\vec r e^{-i\vec{\tilde p}\cdot \vec r} \phi_{\vec k}(\vec r)\ov k^2-p^2-i\ep} = 
\delta(\vec p-\vec k) \\ \nonumber
+{1\ov (2\pi)^3}\lim_{\vec{\tilde p}-\vec p}( {\tilde p}^2-p^2)
{\int d\vec r e^{-i\vec{\tilde p}\cdot \vec r} \eta_{\vec k}(\vec r)\ov k^2-p^2-i\ep} 
\eea

The second term in the r.h.s. is zero because 
${\int d\vec r e^{-i\vec p\cdot \vec r} \eta_{\vec k}(\vec r)\ov k^2-p^2-i\ep} $ is finite
(and also its integral over $\vec k$ is finite). 
We limit ourselves to check the first order in perturbation theory:
\be
\eta_{\vec k}(\vec r)={1\ov -\p^2-k^2+i\ep}\cdot {2m_r\alpha e^{-\mu r}\ov r} e^{i\vec k\cdot\vec r}=
4\pi^2 m_r\alpha\int d\vec q e^{i\vec q\cdot\vec r}{1\ov q^2-k^2+i\ep}
{1\ov (\vec q-\vec k)^2+\mu^2}
\ee
therefore
\be
{1\ov (2\pi)^3}{\int d\vec r e^{-i\vec p\cdot \vec r} \eta_{\vec k}(\vec r)\ov  k^2-p^2-i\ep}=
-{4\pi^2 m_r\alpha\ov (p^2-k^2+i\ep)^2}{1\ov (\vec p-\vec k)^2+\mu^2}
\ee
This is finite for generic $p$ and $k$. Moreover its integral over $\vec k$ is also finite:
by first doing the angular integration we get
\be
{1\ov (2\pi)^3}\int d\vec k{\int d\vec r e^{-i\vec p\cdot \vec r} \eta_{\vec k}(\vec r)\ov k^2-p^2-i\ep}=
-{2\pi^3 m_r\alpha\ov p}
\int_{-\infty}^{+\infty}dk k {log{(k+p)^2+\mu^2\ov (k-p)^2+\mu^2}\ov (k^2-p^2-i\ep)^2}
\ee
This is seen to be finite by closing the contour in the upper plane, the singularities being 
a double pole at $k=p+i\ep$ and branch cuts at $k=\pm p+i\mu$.

Therefore there are no hidden $\delta$ functions:
\be
\int d\vec k{1\ov (2\pi)^3}\lim_{\vec{\tilde p}-\vec p}( {\tilde p}^2-p^2)
{\int d\vec r e^{-i\vec{\tilde p}\cdot \vec r} \eta_{\vec k}(\vec r)\ov k^2-p^2-i\ep} =0
\ee

 {\bf Acknowledgments} It is a pleasure to aknowldge Marco Serone for drawing my attention
to the Sommerfeld effect and for asking relevant questions, Piero Ullio for discussions and for pointing  the relevance of the P wave, and Serguey Petcov for encouragements.


\begin{thebibliography}{99}
\bibitem{Hisa}
J.Hisano,S.Matsumoto,M.M.Nojiri,O.Saito, 
"Non-Perturbative Effect on Dark Matter Annihilation and
Gamma Ray Signature from Galactic Center" 
Phys.Rev.D71,063528 (2005)
\bibitem{Stru}
M.Cirelli,A.Strumia,M.Tamburini,
"Cosmology and Astrophysics of Minimal Dark Matter"
Nucl.Phys. B800:204 (2008), 
arXiv:0706.4071 [hep-th]
\bibitem{Mar}
J.March-Russel,S.M.West,D.Cumberba,D.Hooper,
"Heavy Dark Matter Through the Higgs Portal"
JHEP0807:058 (2008), 
ArXiv: 0801.3440 [hep-ph]]
\bibitem{Stru2}
A.Strumia,
"Sommerfeld corrections to type-II and III leptogenesis"
Nucl.Phys.B809:308 (2009), 
arXiv:0806.163 [hep-ph]
\bibitem{ArkHam}
N.Arkani-Hamed,D.P.Finkbeiner,T.Slatyer,N.Weiner,
"A Theory of Dark Matter"
Phys.Rev. D79:015014 (2009), 
arXiv:0810.071 [hep-ph]
\bibitem{Lat}
M.Lattanzi,J.Silk,
"Can the WIMP annihilation boost factor be boosted by the Sommerfeld enhancement?"
arXiv: 0812.0360 [astro-ph]
\bibitem{Cli}
F.Chen,J.M.Cline,A.R.Frey, 
"A new twist on excited dark matter: implications for INTEGRAL, PAMELA/ATIC/PPB-BETS,DAMA" 
arXiv:0901.4327 [hep-ph] 
\bibitem{Rob}
B.E.Robertson, A.R.Zentner,
"Dark matter annihilation rates with velocity-dependent annihilation cross section"
arXiv:0902.0362 [astro-ph]
\bibitem{LL}
L.D.Landau,E.M.Lifshitz,
"Quantum Mechanics" 
Pergamon Press 1958
\bibitem{IZ}
C.Itzykson,J-B.Zuber, 
"Quantum Field Theory"
McGraw-Hill 1985
\bibitem{Pesk}
M.E.Peskin,D.V.Schroeder,
"An Introduction to Quantum Field Theory" Perseus Books 1995

\end{thebibliography}
\end{document}